\documentclass[12pt]{article}

\usepackage{putex}
\usepackage{graphicx}
\usepackage{latexsym,amsmath,amsfonts,amssymb}
\usepackage{bbm}

\def\no{\nonumber \\}
\def\btab{\begin{table}[h] \begin{center} \begin{tabular}{l lp{3in}}}
      \def\etab{\end{tabular} \end{center} \end{table}}
\def\btabm{\begin{center} \begin{tabular}}
    \def\etabm{\end{tabular} \end{center}}

\def\ie{{\it i.e.}}

\def\m{{\mu}}
\def\n{{\nu}}

\def\s{\sqrt}

\def\CE{{\cal E}}

\def\CN{{\cal N}}
\def\CO{{\cal O}}
\def\CP{{\cal P}}
\def\CQ{{\cal Q}}
\def\CW{{\cal W}}

\def\CS{{\cal S}}

\begin{document}

\title{Flows to Schr{\" o}dinger Geometries}

\authors{Takaaki Ishii$^\diamondsuit$, and Tatsuma Nishioka$^\spadesuit$}

\institution{DAMPT}{
${}^\diamondsuit$Department of Applied Mathematics and Theoretical Physics, \cr
University of Cambridge, Cambridge CB3 0WA, UK}
\institution{PU}{
${}^\spadesuit$Department of Physics, Princeton University, Princeton, NJ 08544, USA}

\abstract{
We construct RG flow solutions interpolating AdS and Schr\"{o}dinger geometries
in Abelian Higgs models obtained from consistent reductions of type IIB supergravity and  M-theory.
We find that $z=2$ Schr\"{o}dinger geometries can be realized at the minima of scalar potentials of these models,
where a scalar charged under $U(1)$ gauge symmetry obtains a nonzero vacuum expectation value.
The RG flows are induced by an operator deformation of the dual CFT.
The flows are captured by fake superpotentials of the theories.
}

\preprint{DAMTP-2011-79 \\ PUPT-2392}

\date{September 2011}

\maketitle

\tableofcontents

\section{Introduction}
Since the discovery of the AdS/CFT correspondence \cite{Maldacena:1997re,Gubser:1998bc,Witten:1998qj}, 
there have been a lot of attempts to apply the correspondence to describe strongly coupled quantum field theories in terms of classical gravity.
After the discovery, gravity duals of relativistic theories were considered at first.
Now it also becomes important to consider applications to strongly coupled nonrelativistic systems.
Gravity duals for nonrelativistic conformal field theories were constructed in
Einstein gravity coupled to massive vector fields which break the relativistic conformal group
to a nonrelativistic one \cite{Son:2008ye,Balasubramanian:2008dm}.
Soon after that, they were embedded into string/M theories using the so-called null Melvin twist solution generating technique
\cite{Maldacena:2008wh,Herzog:2008wg,Adams:2008wt}.
These solutions are called Schr\"{o}dinger geometry as the geometries have Schr\"{o}dinger symmetry.
Lifshitz geometry is also important for considering the gravity dual of nonrelativistic field theories,
and is also obtained as a solution in the massive gauge theory \cite{Kachru:2008yh,Taylor:2008tg},
which can be obtained from the Abelian Higgs model \cite{Gubser:2009cg}. 
The embeddings to the string/M theories are worked out in \cite{Balasubramanian:2010uk,Donos:2010tu,Gregory:2010gx,Donos:2010ax,Cassani:2011sv,Halmagyi:2011xh}.
But the embedding of the Lifshitz geometries into string/M theory is more involved than that of the Schr\"{o}dinger ones.

In this paper, we will consider a realization of the Schr{\" o}dinger geometry in the Abelian Higgs models given by
\cite{Gubser:2009qm,Gauntlett:2009dn}, which are consistent truncations of type IIB SUGRA and M-theory.\footnote{
More general reductions, including these, are 
considered in \cite{Gauntlett:2010vu,Cassani:2010uw}.}
These models have W-shaped scalar potentials with two extrema at the origin and the nonzero point of a charged scalar field.
The Abelian Higgs model is a massive gauge theory around the nonzero point, where 
the expectation value of the scalar field gives rise to the mass of the gauge field.
Therefore, the Schr\"{o}dinger geometry is expected to be realized there, while
the asymptotically AdS space is obtained at the origin of the scalar potentials.
It is, however, still nontrivial to embed the nonrelativistic geometries into these models since 
the parameters in the scalar potentials cannot be arbitrarily varied as opposed to effectively constructed bottom-up models.
We will see that it is possible to find the Schr{\" o}dinger solutions in both cases, while it is not possible for the Lifshitz solutions.

In particular, since the AdS and the Schr\"{o}dinger geometries are realized in the maximum and the minimum of the potential, respectively, 
there must be an interpolating solution between them.
We will construct such a solution, which is a domain-wall dual to the RG flow
from a relativistic CFT in the ultraviolet (UV) to a nonrelativistic CFT in the infrared (IR).\footnote{
In a different context, flow solutions from $AdS_4$ spacetime to the five-dimensional Schr{\" o}dinger geometry
have been constructed in M-theory in \cite{Kim:2011fb}, whose IR geometry was obtained in \cite{O'Colgain:2009yd}.
See also \cite{Ooguri:2009cv,Jeong:2009aa} for some related works on the geometric realization of Schr\"{o}dinger symmetry. }
To trigger the RG flow, we will consider  an addition of the deformation term $J \CO$ to the UV theory,
where $J$ is an external field coupled to the scalar operator $\CO$ charged under a $U(1)$ symmetry.
With this deformation, the $U(1)$ symmetry is supposed to be explicitly broken,
and it is expected to flow to the other IR fixed point, where the IR geometry is the Schr\"{o}dinger geometry in our case.

This paper is organized as follows. In Sec.\,2, we review how to embed 
the Schr{\" o}dinger geometry into the Abelian Higgs model,
and then we consider the string/M theory construction of the Schr\"{o}dinger geometries.
In Sec.\,3, we will construct the interpolating solutions. After discussing the asymptotics of the fields,
we obtain domain-wall solutions numerically.
We will discuss the puzzling aspects of our solutions and possible future directions in Sec.\,4.

\section{Schr{\" o}dinger geometries from Abelian Higgs model}
Gravity duals for nonrelativistic conformal field theories can be constructed in
Einstein gravity coupled to massive vector fields \cite{Son:2008ye,Balasubramanian:2008dm}.
It is also pointed out that such massive gauge theories can be realized
in the broken phase of Abelian Higgs models \cite{Balasubramanian:2008dm}.
Let us summarize the construction briefly.

The metric of a gravity background dual to the boundary theories with Schr\"{o}dinger symmetry is given by \cite{Son:2008ye,Balasubramanian:2008dm}
\begin{align}\label{Sh}
	ds^2 = \ell^2 \left[ -\sigma^2 r^{2z} (dx^+)^2 + \frac{dr^2}{r^2}
	+ r^2 (-dx^+ dx^- + d\vec x^2) \right] \ ,
\end{align}
where $z$ is the dynamical exponent, $\ell$ the radius, and $\sigma$ a parameter of the geometry.
This arises from Einstein gravity coupled to a massive gauge field,
\begin{align}
	S = \frac{1}{16\pi G_N} \int d^{d+3}x \s{-g} \left( R - 2\Lambda - \frac{1}{4}F_{\m\n}F^{\m\n}
	-\frac{m^2}{2} A_\m A^\m \right) \ .
\end{align}
The solution \eqref{Sh} exists provided that 
the cosmological constant and the mass of the vector field are chosen as
\begin{align}\label{SchPar}
	\Lambda = -\frac{(d+1)(d+2)}{2\ell^2} \  , \qquad m^2 = \frac{z(z+d)}{\ell^2} \ ,
\end{align}
and that the gauge field takes the form
\begin{align}
	A_+ = 2\sigma\ell \s{\frac{z^2 -1}{z^2 +8}}~ r^z \ .
\end{align}

The massive gauge theory can be obtained in the broken phase of the Abelian Higgs model 
\begin{align}\label{AHM}
S= \frac{1}{16\pi G_N} \int d^{d+3}x \s{-g} \left( R - \frac{1}{4}F_{\m\n}F^{\m\n}
	-|(\partial_\m - i q A_\m)\psi|^2 - V(\psi)
 \right) \ ,
\end{align}
in which the scalar field $\psi$ obtains a nonzero vacuum expectation value (VEV) in the minima of the scalar potential $V(\psi)$.
The form of $\psi$ is restricted as $\psi = \psi_0 e^{i\theta}$,
where $\psi_0 \in \mathbb{R}$ corresponds to the vacuum expectation value of the scalar field.
The phase $\theta$ can be absorbed by the gauge shift, and
the resulting theory is the massive gauge theory,
\begin{align}
	S= \frac{1}{16\pi G_N} \int d^{d+3}x \s{-g} \left( R - \frac{1}{4}F_{\m\n}F^{\m\n}
	- q^2 \psi_0^2 A_\mu A^\mu - V(\psi_0^2) 
 \right) \ .
\end{align}
It follows that the Schr\"{o}dinger geometry can be obtained in the Abelian Higgs model \eqref{AHM}
with the following parameters \cite{Balasubramanian:2008dm},
\begin{align}
	q^2 \psi_0^2 = \frac{z(z+d)}{2\ell^2} \ , \qquad
	V(\psi_0^2 ) = -\frac{(d+1)(d+2)}{\ell^2} \ .
\end{align}

We will consider the Schr\"{o}dinger geometries in the models obtained from consistent truncations of
ten-dimensional type IIB SUGRA and eleven-dimensional M-theory to five and four dimensions, respectively.\footnote{
More general solutions are constructed in \cite{Donos:2009xc}, where the internal spaces are taken as Sasaki-Einstein manifolds,
while they are squashed in our case.}
In both cases, we will obtain the solutions with the dynamical exponent $z=2$.

\subsection{In IIB string theory}\label{ss:Sch_IIB}

Let us start from the five-dimensional case of type IIB SUGRA. The action contains
a metric, a $U(1)$ gauge field, and a scalar \cite{Gubser:2009qm},\footnote{
Here we rescaled the gauge field $A_{ours} = \frac{2L}{\s{3}}A_{theirs}$ to normalize the coefficient of the field strength.
}
\begin{align}\label{KK_IIB}
	S= \frac{1}{16\pi G_N} &\int d^5 x \s{-g} \Big[ R - \frac{1}{4}F^2 + \frac{1}{12\s{3}}
	\epsilon^{\lambda\m\n\sigma\rho}F_{\lambda\m}F_{\n\sigma}A_\rho \no
	&\qquad -\frac{1}{2}(\partial_\m \eta)^2 - \frac{\sinh^2 \eta}{2}(\partial_\m \theta 
	-\frac{\s{3}}{L}A_\m)^2 + \frac{3}{L^2}\cosh^2\frac{\eta}{2}(5-\cosh\eta) \Big] \ .
\end{align}
This action is derived from a consistent truncation of a ten-dimensional IIB SUGRA action
by decomposing the metric as
\begin{align}
	ds^2 = \cosh\frac{\eta}{2} ds_5^2 + \frac{L^2}{\cosh\frac{\eta}{2}} \left[ 
	ds_V^2 + \cosh^2\frac{\eta}{2} (\zeta^A)^2\right] \ ,
\end{align}
where $V$ is a four-dimensional K{\" a}hler-Einstein manifold with $R_{\m\n} = 6g_{\m\n}$,
and $\zeta^A = \zeta + \frac{A}{\s{3}L}$ and $\zeta = d\psi + \sigma$ such that $d\zeta = 2\omega$, with
$\omega$ being a K{\" a}hler form \cite{Gubser:2009qm}.
The phase $\theta$ can be absorbed by the gauge shift.
The Chern-Simons term can be ignored when the gauge field is assumed as $A \sim dx^+$.
The potential minima for the scalar field $\eta$ are at
\begin{align}\label{etamin_IIB}
	\eta = \pm \log (2+\s{3}) \ .
\end{align}
The positive sign in \eqref{etamin_IIB} is chosen without loss of generality.
The action reduces to the massive gauge theory around the minima,
where the radius, the cosmological constant, and the mass of the gauge field are read off as
\begin{align}
	\ell^2 = \frac{8}{9} L^2 \ , \qquad \Lambda = -\frac{27}{4L^2} \ , \qquad m^2 = \frac{9}{L^2} \ .
\end{align}
Comparing them with \eqref{SchPar}, the $z=2$ Schr{\" o}dinger geometry is obtained
in this model.\footnote{
There exists the other solution of \eqref{SchPar} with $z=-4$. We do not adopt it here.}
This solution is not supersymmetric because the $U(1)$ fibration of the internal manifold is stretched
and there is no Killing spinor on it \cite{Romans:1984an}, as is mentioned in \cite{Gubser:2009qm} for
the AdS solution.

\subsection{In M-theory}\label{ss:Sch_M}

Similarly we can use a four-dimensional theory obtained from a consistent truncation 
of M-theory on a seven-dimensional Sasaki-Einstein manifold \cite{Gauntlett:2009dn,Gubser:2009gp},
\begin{align}\label{KK_M}
	S= \frac{1}{16\pi G_N} &\int d^4 x\s{-g} \Big[  R - \frac{1}{4} F^2 -\frac{1}{2}(\partial_\m \eta)^2 \no
	&\qquad  - \frac{\sinh^2 \eta}{2}(\partial_\m \theta 
	-\frac{1}{L}A_\m)^2 + \frac{1}{L^2}\cosh^2\frac{\eta}{2}(7-\cosh\eta)
	 \Big] \ .
\end{align}
Here we need a constraint for the gauge field $F \wedge F = 0$, but
this is always satisfied for the Schr{\" o}dinger background.
The potential minima are at 
\begin{align}
\eta  = \pm\log (3 + 2\s{2}) \ ,
\end{align}
and we find the massive vector theory
with the parameters 
\begin{align}\label{Mpar}
	\ell^2 = \frac{3}{4}L^2 \ , \qquad \Lambda = -\frac{4}{L^2} \ , \qquad m^2 = \frac{8}{L^2} \ .
\end{align}
Then we obtain $z = 2$ by solving \eqref{SchPar} and \eqref{Mpar}.\footnote{$z=-3$ is the other solution which satisfies \eqref{SchPar}.} This solution is also nonsupersymmetric for a reason similar to the type IIB case \cite{Pope:1984jj,Pope:1984bd}.

\subsection{Comment on Lifshitz solutions}
The massive vector theory also allows us to construct Lifshitz solutions of the form
\cite{Kachru:2008yh,Taylor:2008tg}
\begin{align}
	ds^2 = - \left(\frac{r}{\ell}\right)^{2z} dt^2 +\ell^2 \frac{dr^2}{r^2} +
	 \left(\frac{r}{\ell}\right)^{2} d{\vec x_{d+1}}^2 \ ,
\end{align}
by tuning the parameters such that
\begin{align}\label{LifPar}
	\Lambda = -\frac{z^2 + d z + (d+1)^2}{2\ell^2} \ , \qquad m^2 = \frac{2z}{\ell^2} \ .
\end{align}
However, there are no physical solutions in the four- and five-dimensional
theories considered above.
It would be interesting to look for a consistent truncation which affords parameters 
consistent with the relations \eqref{LifPar}.

\section{Domain-wall solutions}
The actions given by \eqref{KK_IIB} and \eqref{KK_M} in five and four dimensions, respectively,
have two vacua: one is supersymmetric at $\eta=0$, 
and the other is nonsupersymmetric at $\eta=\eta_\ast\neq0$.
As already studied in \cite{Cassani:2010uw,Gauntlett:2009zw},
domain-wall solutions interpolating two AdS vacua can be constructed.
Here we expect a domain-wall solution interpolating the AdS and the Schr{\" o}dinger geometries.

We assume the following ansatz for the domain-wall solution,
\begin{align}\label{Ansatz}
	ds^2  = &L^2 \left( d\rho^2 + e^{2A(\rho)}((d{x^i})^2 - dx^+ dx^-) - B(\rho) e^{2zA(\rho)} (dx^+)^2 \right) \ , \notag\\
	 A_+ &= L a(\rho) e^{zA(\rho)} \ , \\
	 \eta &= \eta (\rho) \ ,\notag
\end{align}
where $i=1,\dots,d$, and $d=2$ and $d=1$ for type IIB and M-theory, respectively.
This ansatz is chosen such that the Schr\"{o}dinger geometry in the IR will become manifest.
We focus only on the case of zero temperature, and an operator deformation will be assumed in the UV.
The equations of motion are given by 
\begin{align}\label{FSG}
	 \eta'' + (d+2)A' \eta' - \partial_\eta V_{pot} &= 0 \ , \nonumber\\
	 2(d+1)A'' + (d+1)(d+2)A'^2 + \frac{1}{2}\eta'^2 + V_{pot} & = 0 \ , \\
	 (d+1)(d+2)A'^2 - \frac{1}{2}\eta'^2 + V_{pot}  & = 0 \nonumber\ , 
\end{align}
where $V_{pot}$ is a scalar potential for $\eta$, 
\begin{align}\label{eomB}
	B''+ (4z+d-2)A'B' - \Big[ 2(d+2-z)A'' - 2(2z^2 +(d-2)z -\frac{d^2+5d+2}{2})A'^2 + \frac{1}{2}&\eta'^2 + V_{pot} \Big] B  \nonumber\\
		= \frac{d(d+1)}{2}a^2 \sinh^2\eta  + (z a A' &+ a')^2 \ ,
\end{align}
and the Maxwell equation is
\begin{align}\label{IIBME}
	a'' + (2z+d)A'a' + \left[ zA'' +z(z+d) A'^2 - \frac{d(d+1)}{2}\sinh^2\eta \right] a = 0 \ .
\end{align}

The second-order differential equations in \eqref{FSG} can be written as first-order equations \cite{Skenderis:1999mm}.
One of the equations \eqref{FSG} is not independent, and
the two independent ones can be obtained as the Euler-Lagrange equations for 
the following functional:
\begin{align}
	E = \int_{-\infty}^\infty d\rho \, e^{(d+2)A(\rho)} \left[ (\partial_\rho \eta)^2 - 2(d+1)(d+2)(\partial_\rho A)^2 + 2V_{pot} \right] \ .
\end{align}
When the potential for the scalar takes the form
\begin{align}\label{FakeSP}
	V_{pot} =  \frac{1}{2} (\partial_\eta \CW)^2 - \frac{d+2}{4(d+1)} \CW^2  \ ,
\end{align}
the functional can be rewritten as 
\begin{align}
	E = \int_{-\infty}^\infty d\rho \, &e^{4A(\rho)} \left[ (\eta'  \mp \partial_\eta \CW)^2 
		- 2(d+1)(d+2) (A' \pm \frac{1}{2(d+1)}\CW)^2 \right] \nonumber \\
		& \pm 2 e^{(d+2)A(\rho)}\CW \big|_{-\infty}^{\infty} \ .
\end{align}
Then it gives two first-order equations of motion,
\begin{align}\label{FSeom_ddim}
	A' = \mp \frac{1}{2(d+1)}\CW \ , \qquad \eta' = \pm  \partial_\eta \CW \ .
\end{align}
Here the $\mathcal{W}$ is called as a ``fake superpotential.''

Let us find out the asymptotic behavior of the fields which solve the equations of motion around $\rho=\pm\infty$.
We first consider the solution in the UV $(\rho \to \infty)$.
The scalar field $\eta$ and metric $A$ can be expanded as
\begin{align}\label{eta_A_asympt}
	\eta &= \eta_{(0)} e^{- \Delta_{-}\rho/L} + \eta_{(1)} e^{-\Delta_{+}\rho/L} + \cdots \ , \nonumber \\
	A &= \frac{\rho}{L} - A_{(0)} e^{-2 \Delta_{-} \rho/L} + \cdots \ ,
\end{align}
where $\Delta_\pm$ are the dimensions of $\eta$ given by the solutions of $\Delta(\Delta-d-2)=- m_\eta^2$,
where $m_\eta$ is the (negative) mass of $\eta$.
Dots in the expansion include further exponentially suppressed contributions.
The presence of $\eta_{(0)}$ corresponds to a deformation of the CFT in the UV by a term $\eta_{(0)} \mathcal{O}$
in the Lagrangian of the boundary theory, and the RG flow is caused by this operator deformation.
The equations of motion \eqref{FSeom_ddim} relates $A_{(0)}$ with $\eta_{(0)}$ as
\begin{align}
	A_{(0)} = \frac{\eta_{(0)}^2}{8(d+1)} \ .
\end{align}
For the Maxwell equation \eqref{IIBME}, we obtain the behavior of the gauge field
\begin{align}
	a = a_{(1)} \, e^{- z \rho/L} + a_{(2)} e^{- (z+d) \rho/L} + \cdots \ ,
\end{align}
which is nothing but
\begin{align}
	A_{+} = L(a_{(1)} + a_{(2)} e^{- d \rho/L} + \cdots) \ .
\end{align}
Finally, the function $B$ takes the form
\begin{align}\label{B_asympt_UV}
	B &= B_{(0)} e^{- m_{-} \rho/L} + B_{(1)} e^{- m_{+} \rho/L} + \cdots \ , \nonumber \\
	m_\pm &= \frac{1}{2} \left[4z+d-2 \pm \sqrt{(4z+d-2)^2 - 8 (2z^2 + (d-2) z - \frac{d^2+5d+2}{2}) + 4 V_{UV} L^2}\right] \ ,
\end{align}
where $V_{UV}$ is the value of the potential at $\rho = \infty$, and here we have used the fact that $\eta^\prime=0$ there.

It is straightforward also to find the asymptotic behavior of the fields in the IR $(\rho \to -\infty)$.
We impose the following boundary condition:
\begin{align}\label{IR_asympt}
	\eta &= \eta_\ast + \widetilde{\eta}_{(0)} e^{\widetilde{\Delta}_{-}\rho/\ell} + \widetilde{\eta}_{(1)} e^{\widetilde{\Delta}_{+}\rho/\ell} + \cdots \ , \nonumber \\
	A &= \frac{\rho}{\ell} - \widetilde{A}_{(0)} e^{2 \widetilde{\Delta}_{-} \rho/\ell} + \cdots \ , \nonumber \\
	a &= a_0 + \cdots \ , \nonumber \\
	B &= a_0^2 + \cdots \ ,
\end{align}
where $\eta_\ast$ is the vacuum expectation value of the scalar field at the minimum of the potential, and 
$\widetilde{\Delta}_\pm$ is the dimension of $\eta$ in the IR.
The fake superpotential $\CW$ defined by \eqref{FakeSP} will be convenient to read off $\widetilde{\Delta}_\pm$.
To derive the asymptotic forms of $a$ and $B$ in \eqref{IR_asympt},
we solved the equations of motion of IIB SUGRA \eqref{KK_IIB} and M-theory \eqref{KK_M} 
around the potential minimum.
The constant $a_0$ appearing in the leading terms of $a$ and $B$ is responsible for the emergence of Schr\"{o}dinger geometry in the IR.
We will numerically solve the equations of motion to find domain-wall solutions with nonzero $a_0$.

\subsection{Fake superpotential and asymptotic solution}

For concreteness, we compute coefficients appearing in the exponents of the asymptotic expansion 
for our IIB SUGRA and M-theory cases. We shall set the AdS radius $L=1$ for simplicity hereafter.
In the computation, it is worthwhile to notice that the fake superpotential is useful to capture the RG flows
and to determine the asymptotics.

\begin{figure}[tb]
	\centering
	\includegraphics[width=7cm]{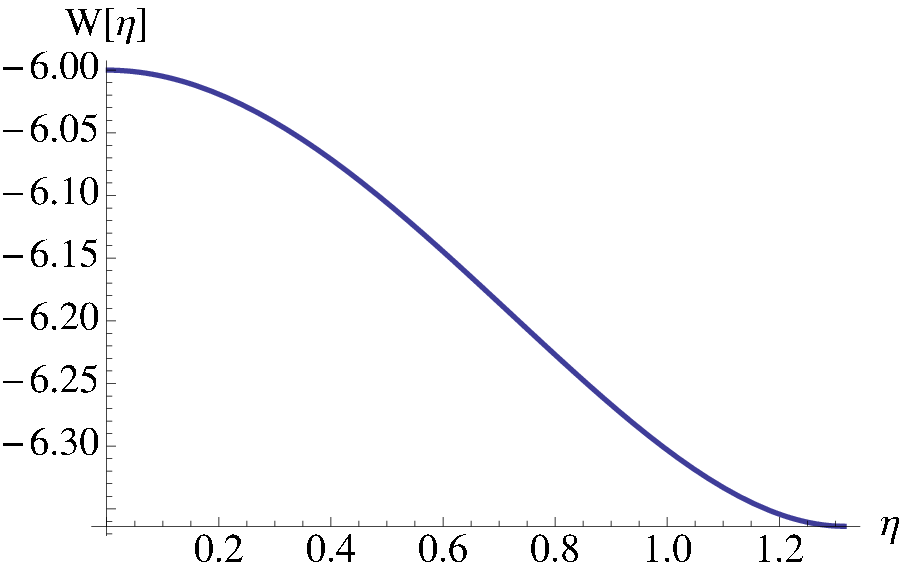}
	\includegraphics[width=7cm]{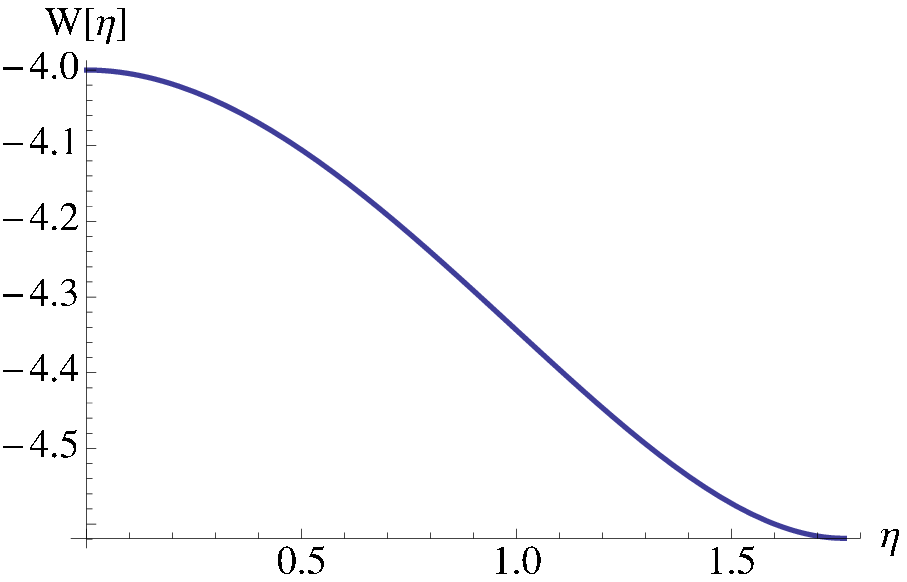}
	\caption{The fake superpotential interpolating supersymmetric $(\eta=0)$ and 
	nonsupersymmetric $(\eta=\eta_\ast)$ vacua in IIB SUGRA (left panel) and M-theory (right panel).}
	\label{fig:fakespot}
\end{figure}

We first consider the IIB SUGRA case, where $d=2$ and the domain-wall solution we focus on has $z=2$. 
We can solve \eqref{FakeSP} to obtain a fake superpotential, as shown in Fig.~\ref{fig:fakespot},
which corresponds to the RG flow induced by the scalar source $\eta_{(0)}$ in the UV and flowing to the vacuum at $\eta=\eta_\ast$ in the IR.
The superpotential can be expanded around $\eta=0$ as $\mathcal{W} = -6 - \eta^2/2 + \cdots$, and we see that
the leading behavior of $\eta$ around $\rho=\infty$ corresponds to $\Delta_{-}=1$.
Since $V_{UV}=-12$ at the UV critical point $\eta=0$, we obtain $m_{-}=2$ from \eqref{B_asympt_UV}.
At the other critical point in the IR, $\eta_\ast = \log (2+\sqrt{3})$, the superpotential in Fig.~\ref{fig:fakespot} can be expanded as
\begin{align}
	\mathcal{W} = -\frac{9}{\sqrt{2}} + \frac{3(\sqrt{6}-\sqrt{2})}{4} (\eta - \eta_*)^2 + \cdots \ ,
\end{align}
and then we obtain $\widetilde{\Delta}_{-}= 2\sqrt{3} - 2$.  

We can repeat this analysis also in the M-theory case, where $d=1$, $z=2$, $V_{UV}=-6$. 
In the UV, we have $\Delta_{-}=1$, and \eqref{B_asympt_UV} gives $m_{-}=2$.
Around the IR critical point, $\eta_\ast = \log (3+2\sqrt{2})$, the superpotential can be expanded as
\begin{align}
	\mathcal{W} = -\frac{8}{\sqrt{3}} + \frac{(\sqrt{11}-\sqrt{3})}{2} (\eta - \eta_*)^2 + \cdots \ ,
\end{align}
and then we obtain $\widetilde{\Delta}_{-}= (\sqrt{33}-3)/2$.

As the asymptotic behavior in the UV is obtained,
it is convenient to rewrite the metric \eqref{Ansatz} so that the AdS geometry is manifest there.
The ansatz \eqref{Ansatz} can be written as
\begin{align}\label{Ansatz_altform}
	ds^2  = L^2 \left[ d\rho^2 + e^{2A(\rho)} \left(- H(\rho) \left(dx^+ + \frac{dx^-}{2 H(\rho)}\right)^2 + \frac{(dx^-)^2}{4 H(\rho)} + (d{x^i})^2 \right) \right] \ ,
\end{align}
where $H(\rho) \equiv B(\rho) e^{2(z-1) A(\rho)}$.
Since $A(\rho) \to r/L$ and $\ B \to B_{(0)} e^{-2r/L}$ as $\rho \to \infty$ for both the IIB SUGRA and the M-theory cases $(z=2)$,
we find $H(\rho) \to  B_{(0)}$ in the UV.\footnote{
It is not suitable to write the metric as in the form \eqref{Ansatz_altform} in the IR because $H(\rho) \to 0$ as $\rho \to -\infty$.
}
Hence the metric in UV is AdS,
\begin{align}
	ds^2  = L^2 \left[ d\rho^2 + e^{2 \rho/L} \left(- d\tau^2 + d\sigma^2 + (d{x^i})^2 \right) \right] \ ,
\end{align}
where $\tau \equiv B_{(0)}^{1/2} x^+ + B_{(0)}^{-1/2} x^-/2$ and $\sigma \equiv B_{(0)}^{-1/2}x^-/2$.
The coefficient $B_{(0)}$, which will be obtained from numerical solutions and should be of order-one,
is related to this ``twisting'' of the AdS geometry to generate the Schr\"{o}dinger geometry.

\subsection{Numerical solutions}
We solve the equations of motion numerically by using the shooting method to obtain domain-wall solutions.
We begin with the case of the IIB SUGRA, and will go back to the M-theory case after that.
We focus only on the leading behavior of the asymptotic behavior of the fields in the UV and the IR.
The subleading terms could be computed with sufficient numerical accuracy.\footnote{
This might be interesting from the point of view of the thermodynamic relation in the Appendix.
}

We have one parameter whose value can be chosen as the initial condition in the UV: $\eta_{(0)}$.
There is a scaling symmetry that allows us to set $a_{(1)}=1$ without loss of generality unless it is zero.
Given this, we can obtain a domain-wall solution where the IR geometry is the Schr\"{o}dinger geometry
whose IR behavior is characterized by the value of $a_0$.
We will show that the $a_0$ is simply determined as a function of $\eta_{(0)}$.

\begin{figure}[tb]
	\centering
	\includegraphics[width=7cm]{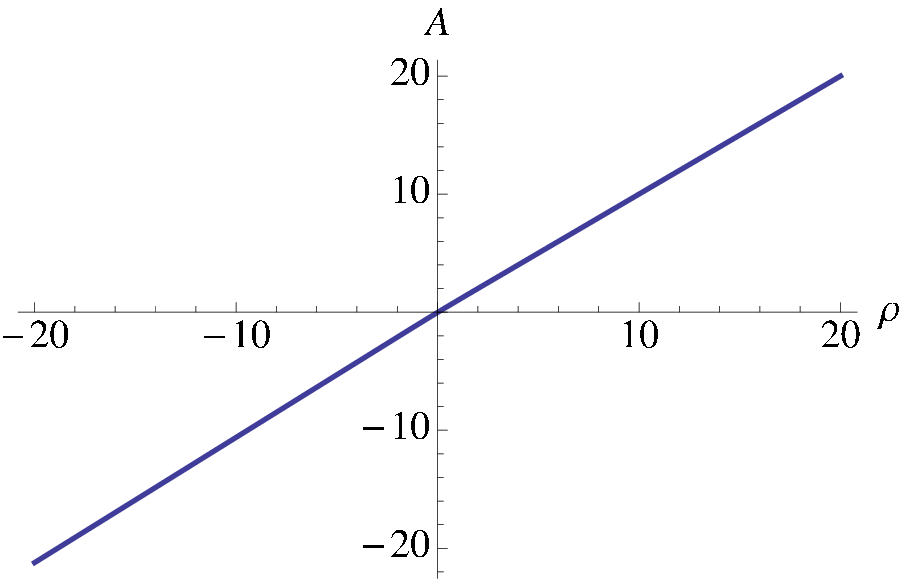}
	\includegraphics[width=7cm]{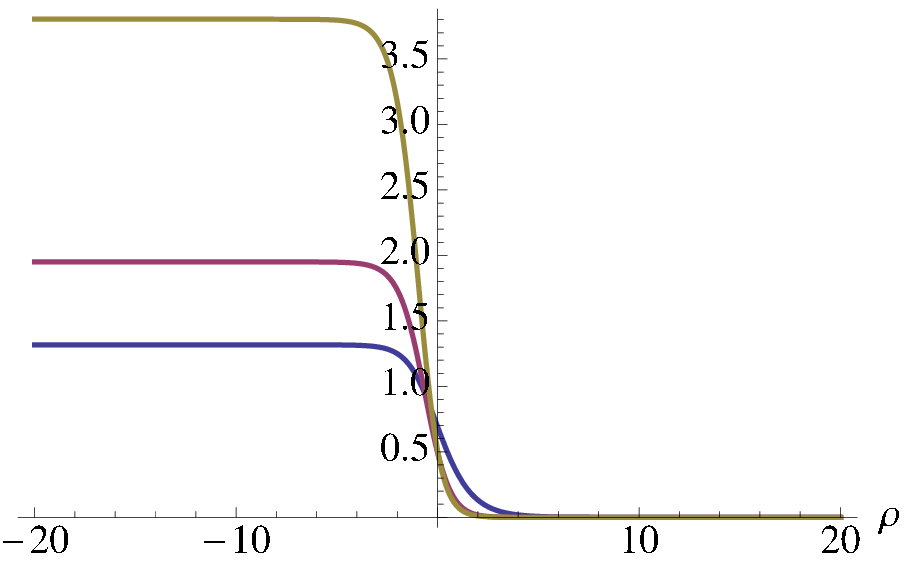}
	\caption{Domain-wall solution interpolating AdS and Schr\"{o}dinger geometries
	in a five-dimensional SUGRA with $\eta_{(0)}=1$. Left panel: The plot of $A(\rho)$.
	Right panel: The blue (bottom), purple (middle), and olive (top) lines correspond to $\eta(\rho)$, $a(\rho)$, and $B(\rho)$, respectively.}
	\label{fig:DW_IIB}
\end{figure}

\begin{figure}[tb]
	\centering
	\includegraphics[width=7cm]{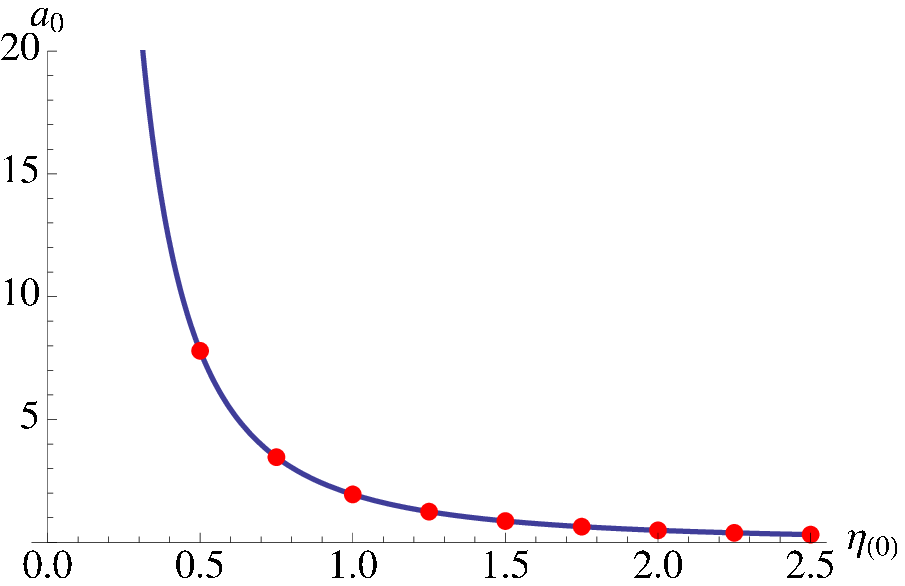}
	\includegraphics[width=7cm]{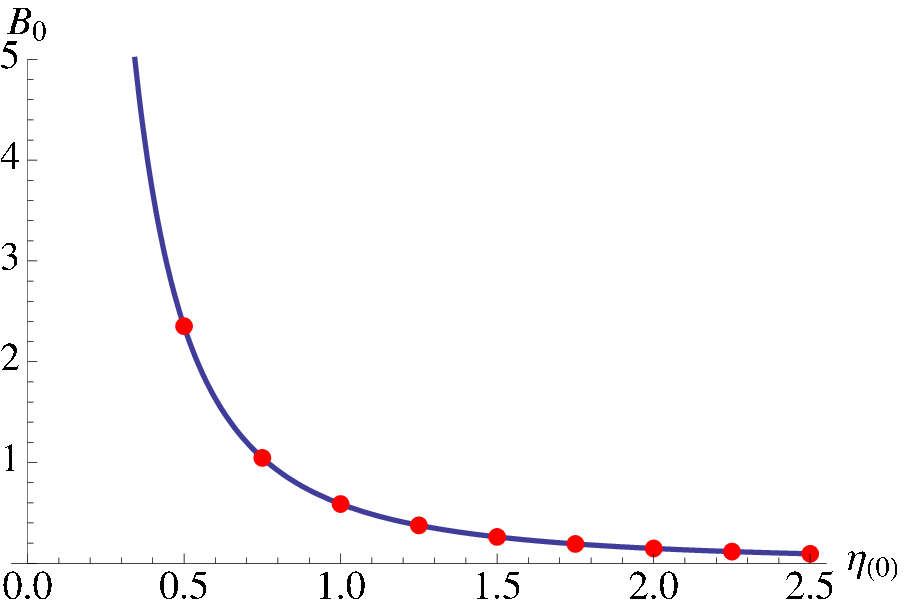}
	\caption{Left panel: The plot of $a_0$ as a function of $\eta_{(0)}$.
	The red dots correspond to the numerical result, and they can be fitted by $1.95 / \eta_{(0)}^2$ as shown by the blue curve.
	Right panel: The plot of $B_{(0)}$ as a function of $\eta_{(0)}$. The plot points can be fitted by $0.588 / \eta_{(0)}^2$.}
	\label{fig:etanu_IIB}
\end{figure}

Figure \ref{fig:DW_IIB} shows the domain-wall solution in the five-dimensional SUGRA \eqref{KK_IIB} with $\eta_{(0)}=1$.
In the IR, the $a_0$ and $B$ grow along with $\eta$, and the domain-wall solution interpolate the AdS in the UV and the Schr\"{o}dinger geometry in the IR.
In the absence of the gauge field, the IR geometry is just another lightlike AdS where the theory is in a nonsupersymmetric vacuum.
The choice of $\eta_{(0)}$ in Fig.~\ref{fig:DW_IIB} is just for demonstration,
so it is interesting to see how $a_0$ behaves when $\eta_{(0)}$ is varied.
The result is shown in Fig.\,\ref{fig:etanu_IIB}, and we find that the plot points can be fitted by $a_0 = 1.95 / \eta_{(0)}^{2}$ and $B_{(0)} = 0.588/\eta_{(0)}^2$.

\begin{figure}[tb]
	\centering
	\includegraphics[width=7cm]{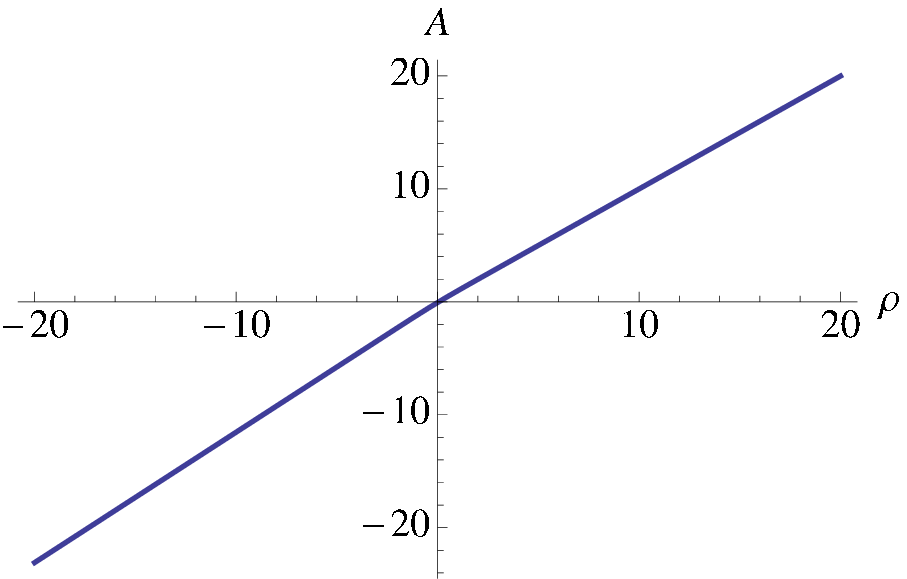}
	\includegraphics[width=7cm]{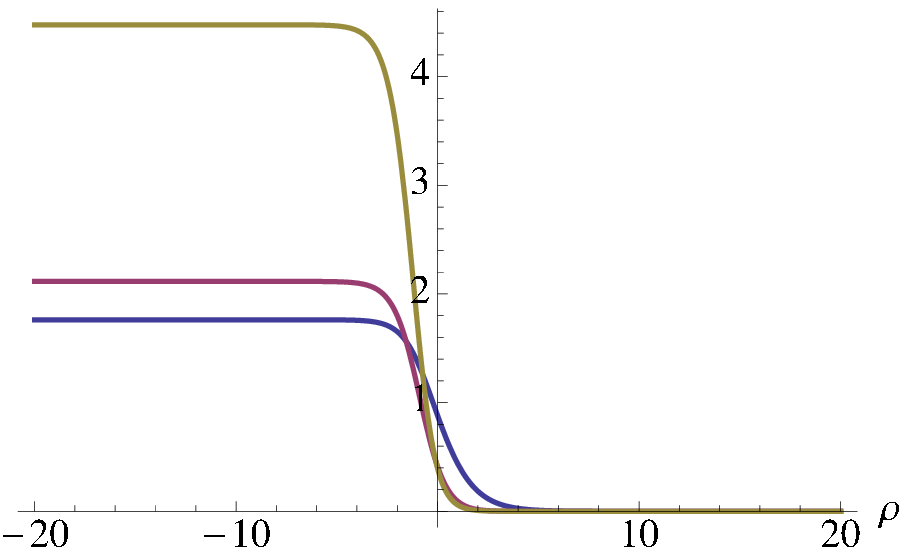}
	\caption{Domain-wall solution interpolating AdS and Schr\"{o}dinger geometries
	in a four-dimensional SUGRA with $\eta_{(0)}=3/2$. 
	Left panel: The plot of $A(\rho)$.
	Right panel: The blue (bottom), purple (middle), and olive (top) lines correspond to $\eta(\rho)$, $a(\rho)$, and $B(\rho)$, respectively.}
	\label{fig:DW_M}
\end{figure}

The domain-wall solution in the case of M-theory can be computed in a similar way as in the case of type IIB SUGRA.
The solution is shown in Fig.~\ref{fig:DW_M}, where $\eta_{(0)}=3/2$ is chosen as a demonstration.
We find that $a_0$ is given as a function of $\eta_{(0)}$, $a_0 = 4.77 / \eta_{(0)}^2$,
and $B_{(0)} = 0.967/\eta_{(0)}^2$ is obtained.

\begin{figure}[tb]
	\centering
	\includegraphics[width=7cm]{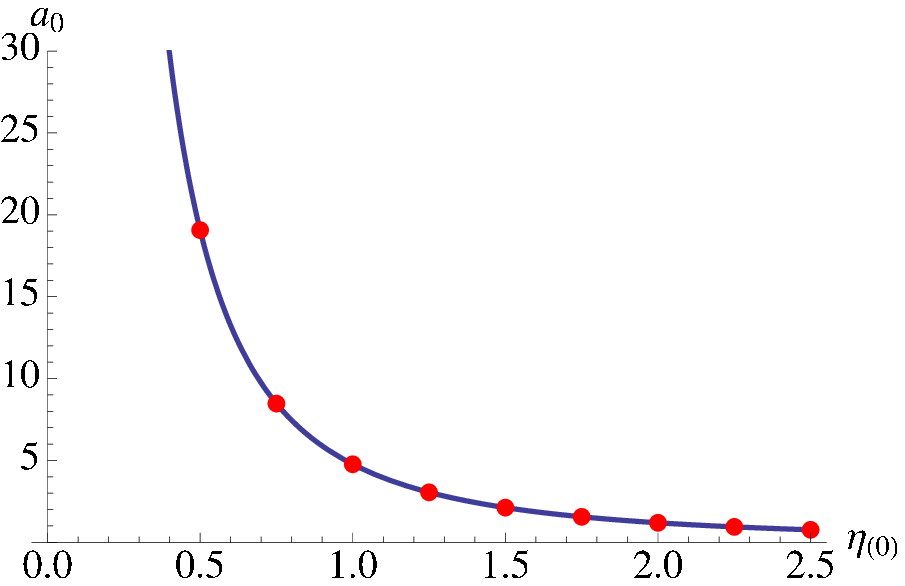}
	\includegraphics[width=7cm]{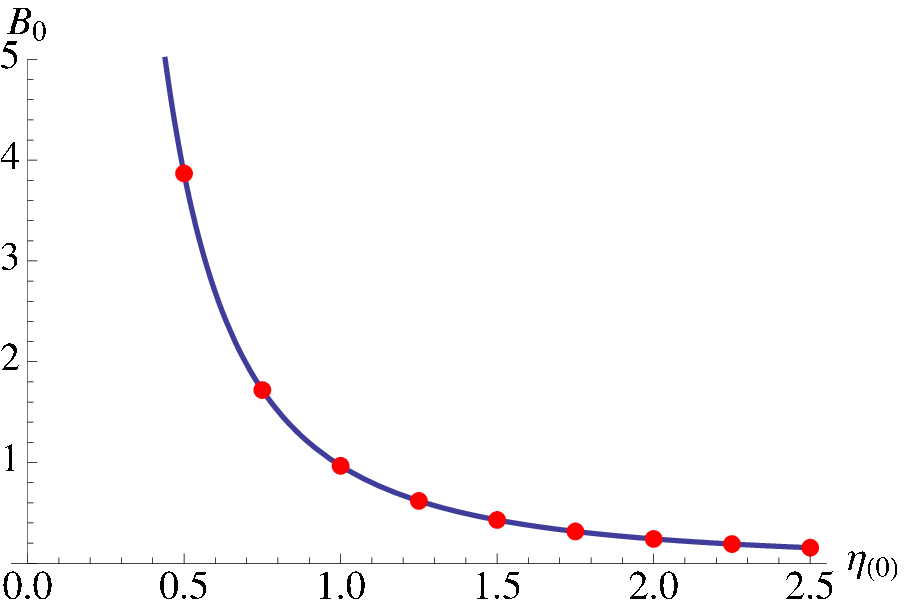}
	\caption{Left panel: The plot of $a_0$ as a function of $\eta_{(0)}$.
	The red dots (numerical results) can be fitted by $4.77 / \eta_{(0)}^2$ as shown by the blue curve.
	Right panel: The plot of $B_{(0)}$ as a function of $\eta_{(0)}$. The plot points can be fitted by $0.967 / \eta_{(0)}^2$.}
	\label{fig:etanu_M}
\end{figure}

The Ricci scalar and the Kretschmann scalar are regular and  monotonic functions of $\rho$ along the RG flow.
In the type IIB SUGRA case, $R=-20/L^2$ and $R_{\mu\nu\rho\sigma} R^{\mu\nu\rho\sigma} = 40/L^4$ in the UV,
and $R=-22.5=-20/\ell^2$ and $RR=50.625=40/\ell^4$ in the IR.
This agrees with the case of AdS-AdS domain-walls.

\section{Discussions}\label{ss:discuss}

We constructed RG flow solutions interpolating AdS and Schr\"{o}dinger geometries
by using Abelian Higgs models obtained in type IIB supergravity and in M-theory.
We found that $z=2$ Schr\"{o}dinger geometries are realized at the minima of the scalar potential of these models,
where the scalar field obtained a nonzero vacuum expectation value,
while it was not possible to embed the Lifshitz geometries into these models.
The Lifshitz geometry has been constructed in four-dimensional $\CN =2$ SUGRA
\cite{Cassani:2011sv,Halmagyi:2011xh} which can be lifted to string/M-theory (see also \cite{Donos:2010tu,Donos:2010ax}).
It would be interesting to investigate consistently truncated Abelian Higgs models 
which allow the Lifshitz geometries as solutions. 

The RG flow was of the type induced by a scalar field source $\eta_{(0)}$ in the UV, corresponding to an operator deformation of the dual CFT.
The solution is a one-parameter family of $\eta_{(0)}$, and the IR value of the massive gauge field was controlled by $\eta_{(0)}$.
What we obtained can be phrased as a realization of emergent nonrelativistic symmetry in the IR at zero temperature.
Our solution, however, is singular when the deformation is turned off, $\eta_{(0)}=0$, as was shown in
Figs.\,\ref{fig:etanu_IIB} and \ref{fig:etanu_M}.
This implies that our RG flow solutions cannot be obtained as a deformation of the AdS-AdS domain walls
of \cite{Gubser:2009gp} despite the fact that both are the solutions of the same theory.
It will be interesting to see how the domain-wall solution we obtained is related to the context of holographic superconductors \cite{Hartnoll:2008vx,Hartnoll:2008kx}.
The domain-wall solution here is induced by the operator deformation in the UV. 
On the other hand, a nonzero chemical potential gives VEVs to the operator dual to the scalar
in the context of holographic superconductors.
It will be challenging to construct the Schr\"{o}dinger geometry by the RG flow induced by the VEV deformation in the UV.
Our solutions would be a special limit of the anisotropic domain walls recently constructed in \cite{Arean:2010wu,Arean:2011gz}, which might provide us with a clear understanding of the dual theories of our solutions.

Since the boundary is the AdS spacetime, we would be able to carry out holographic renormalization
to understand the features of the dual field theory.\footnote{A holographic renormalization
on the Schr{\" o}dinger geometries was discussed in \cite{Guica:2010sw}.} 
Our solutions are similar to the GPPZ flow \cite{Girardello:1999bd}, where the gravity dual of 
an operator deformation of $\CN =4$ SYM was studied.
The holographic renormalization of the GPPZ flow was carefully carried out in \cite{Bianchi:2001kw},
and it is straightforward to repeat the same analysis for our cases.
Incongruously, the $U(1)$ current is conserved despite the operator deformation.
This might be because the expectation value of the $U(1)$ charged operator dual to the source $\eta_{(0)}$
would be zero, \ie, $\eta_{(1)}=0$.
The value of the $\eta_{(1)}$ is completely determined by the behavior of the superpotential around $\eta = 0$.
Although the superpotential we used is numerically generated and does not have a good asymptotic expansion around
$\eta = 0$,  the $\eta_{(1)}$ is likely to be zero within the numerical error. 
Another hint for the existence of the $U(1)$ symmetry is the conserved current we found
in the Appendix that results from the scaling symmetry of our solutions.
The conserved current could be interpreted as the thermodynamic relation of the dual field theory,
and there might be a conserved charge of the form \eqref{TRC} that apparently comes from
the bulk gauge field.
Unfortunately, the usual procedure of holographic renormalization does not give us
such a contribution since the Maxwell action vanishes when the on-shell solution is substituted.
To obtain a nonzero contribution to the on-shell Maxwell action, the $A_-$ component
on the lightlike AdS spacetime will be necessary.
It is not clear how to account for the conservation of the $U(1)$ current (coming from
the bulk gauge symmetry)
as well as  the conserved charge \eqref{TRC} in
the thermodynamic relation [this is not necessarily the same as the $U(1)$ symmetry mentioned above].
Our solutions might be a singular limit of general solutions including the $A_-$ component, so
it would be necessary to relax our ansatz to clarify the property of our solutions.

 \vspace{1.3cm}
 \centerline{\bf Acknowledgements}
We are grateful to K.\,Murata, K.\,Skenderis, Y.\,Tachikawa and M.\,Taylor for valuable discussions.
The work of T.\,N. was supported in part by the U.S.\ NSF under Grants No.\,PHY-0844827 and No.\,PHY-0756966.

\appendix
\section{Conserved charges and thermodynamic relation}\label{ss:thermo}
We would like to evaluate the conserved charges from our solutions that are asymptotically
AdS space. 
We follow the argument given in Appendix C.2 of \cite{Maldacena:2008wh}.
Given the Fefferman-Graham coordinates of the $AdS_{D+1}$ space
\begin{align}
	ds^2_{D+1} = L^2 (d\rho^2 + e^{2\rho} g_{ab}(x,\rho) dx^a dx^b) \ ,
\end{align}
where $g_{ab}$ has the following expansion form
\begin{align}
	g(x,\rho) = g_{(0)} + g_{(2)} e^{-2\rho} + \cdots + g_{(D)} e^{-D\rho} -2 h_{(D)}\rho e^{-D\rho} + \cdots \ ,
\end{align}
the coefficient $h_{(D)}$ is related to the Weyl anomaly of the boundary field theory, and 
the boundary stress-energy tensor reads
\begin{align}
	T_{ab} = \frac{D L}{16\pi G_N} (g_{(D)})_{ab} \ ,
\end{align}
for the flat boundary $(g_{(0)})_{ab} = \eta_{ab}$ \cite{deHaro:2000xn,Skenderis:2000in}.
In our ansatz \eqref{Ansatz}, the metric becomes asymptotically lightlike AdS space by shifting the coordinates as follows:
\begin{align}
	\tilde x^+ = x^+ \ , \qquad \tilde x^- = x^- + B_{(0)}x^+ \ .
\end{align}
The stress-energy tensor vanishes, except for the $++$ component
\begin{align}
	T_{++} = - \frac{(d+2)L}{16\pi G_N} B_{(1)} \ .
\end{align}
Then, the energy density and the particle number density, associated with the Killing vectors $\partial_{\tilde x^+}$
and $\partial_{\tilde x^-}$, respectively, are
\begin{align}\label{EDPND}
	\CE = T_{++} = - \frac{(d+2)L}{16\pi G_N} B_{(1)} \ , \qquad \CN = 0 \ .
\end{align}

One can find the conserved current 
in a similar manner to \cite{Klebanov:2010tj},
\begin{align}\label{Current}
	j = e^{dA} \left( e^{2A}(e^{2(z-1)A} B)' - \phi \phi' \right) \ , \qquad \phi \equiv e^{zA} a \ ,
\end{align}
which satisfies $\partial_\rho j(\rho) = 0$ up to the equations of motion.
It is associated with the scale invariance of the solution of the form \eqref{Ansatz},
\begin{align}
	A  \to  A - \log \lambda \ , \qquad x^i  \to  \lambda x^i \ , \qquad
	 x^+  \to  \lambda^z x^+ \ , \qquad x^-  \to  \lambda^{2-z} x^- \ .
\end{align}
The conservation of the current gives rise to the relation between the parameters in the UV and IR regions,
\begin{align}
	j(\rho= \infty) = j(\rho = -\infty) \ .
\end{align}
This gives us an additional constraint for the parameters as follows:
\begin{align}\label{param_relation}
	\frac{d+2}{d}B_{(1)} =  a_{(1)} a_{(2)} \ .
\end{align}
In the nonrelativistic scale-invariant theory, the energy density and pressure
are related by
\begin{align}
	z \CE = d \CP \ .
\end{align}
Then, the thermodynamic relation gives rise to
\begin{align}
	\frac{d+z}{d}\CE =  T\CS + \mu \CQ \ ,
\end{align}
where $T$ and $\CS$ are the temperature and entropy density, and 
$\mu$ and $\CQ$ are the chemical potential and charge density associated with
some symmetry, respectively.
Since there is no temperature in our solutions and $B_{(1)}$ is proportional to the energy density $\CE$,
we obtain the following relation
\begin{align}\label{TRC}
	\mu\CQ = - \frac{(d+2)L}{16\pi G_N}a_{(1)}a_{(2)} \ .
\end{align}
This implies the existence of a conserved charge in our solutions, but the $U(1)$ symmetry
is supposed to be explicitly broken by the operator deformation. 
We have the particle number symmetry in our solutions, and the density is zero 
as was given in \eqref{EDPND}.
Then there seems to be no charges corresponding to \eqref{TRC}.

This thermodynamic relation is unchanged in the $(\tau, \sigma)$-coordinate.
The components of the gauge field in this coordinate can be identified as
$A_\tau = B_{(0)}^{-1/2} A_+$ and $A_\sigma = -B_{(0)}^{-1/2} A_+$ in $\rho \to \infty$,
and hence the asymptotic behavior of $A_\tau$ is
\begin{align}
	A_\tau = L (a_{(1)} B_{(0)}^{-1/2} + a_{(2)} B_{(0)}^{-1/2} e^{-d \rho/L} + \cdots) \ .
\end{align}
On the other hand, the expansion of the metric gives $g_{(D)\tau\tau}=B_{(1)}/B_{(0)}$.
Therefore, \eqref{param_relation} is not disrupted by the coordinate transformation.
In the $(\tau, \sigma)$ coordinates, the situation is  that the $A_\sigma$ as well as the $A_\tau$ are turned on in the UV.


\bibliographystyle{JHEP}
\bibliography{hscSch_v2}

\end{document}